\documentclass[conference]{IEEEtran}
\IEEEoverridecommandlockouts
\usepackage{cite}
\usepackage{amsmath,amssymb,amsfonts}
\usepackage{algorithmic}
\usepackage{pdflscape}
\usepackage{graphicx}
\usepackage{setspace}
\usepackage{tabu}
\usepackage{textcomp}
\usepackage{xcolor}
\usepackage[inline]{enumitem}
\def\BibTeX{{\rm B\kern-.05em{\sc i\kern-.025em b}\kern-.08em
    T\kern-.1667em\lower.7ex\hbox{E}\kern-.125emX}}

\newcommand{\vtheta}{\ensuremath{\mathrm{\boldsymbol{\theta}}}}
\newcommand{\vx}{\ensuremath{\mathrm{\bold{x}}}}

\allowdisplaybreaks

\newcommand{\vomega}{\ensuremath{\mathrm{\boldsymbol{\omega}}}}

\newcommand{\vnet}{\ensuremath{\mathrm{\bold{f}}}}

\begin{document}
\bstctlcite{IEEEexample:BSTcontrol}

\title{
NeuSpin: Design of a Reliable Edge Neuromorphic System Based on Spintronics for Green AI
}

\author{\IEEEauthorblockN{Soyed Tuhin Ahmed\IEEEauthorrefmark{9}\IEEEauthorrefmark{2}, Kamal Danouchi\IEEEauthorrefmark{3}, 
Guillaume Prenat\IEEEauthorrefmark{3}, Lorena Anghel\IEEEauthorrefmark{3}, Mehdi B. Tahoori\IEEEauthorrefmark{2}\\}
\IEEEauthorblockA{\IEEEauthorrefmark{2}Karlsruhe Institute of Technology, Karlsruhe, Germany, \IEEEauthorrefmark{9}corresponding author, email: soyed.ahmed@kit.edu}
\IEEEauthorblockA{\IEEEauthorrefmark{3}Univ. Grenoble Alpes, CEA, CNRS, Grenoble INP, and IRIG-Spintec, Grenoble, France}
}

\maketitle
\addtolength\abovedisplayskip{-0.6em}%
\addtolength\belowdisplayskip{-0.6em}%
\setlength{\textfloatsep}{0pt}
\setlength\belowdisplayskip{0pt}

\begin{abstract}

Internet of Things (IoT) and smart wearable devices for personalized healthcare will require storing and computing ever-increasing amounts of data. The key requirements for these devices are ultra-low-power, high-processing capabilities, autonomy at low cost, as well as reliability and accuracy to enable Green AI at the edge.
Artificial Intelligence (AI) models, especially Bayesian Neural Networks (BayNNs) are resource-intensive and face challenges with traditional computing architectures due to the memory wall problem. Computing-in-Memory (CIM) with emerging resistive memories offers a solution by combining memory blocks and computing units for higher efficiency and lower power consumption. However, implementing BayNNs on CIM hardware, particularly with spintronic technologies, presents technical challenges due to variability and manufacturing defects. The NeuSPIN project aims to address these challenges through full-stack hardware and software co-design, developing novel algorithmic and circuit design approaches to enhance the performance, energy-efficiency and robustness  of BayNNs on sprintronic-based CIM platforms.

\end{abstract}


\section{Introduction}

Artificial Intelligence's utilization of cloud computing presently accounts for 76\% of the overall traffic in data centers~\cite{gupta2021chasing}. Projections indicate a fivefold surge of this percentage from 2016 to 2021. By 2025, it is anticipated to constitute nearly 20\% of the world's electricity consumption~\cite{magazzino2021investigating}. The emergence of new applications in personalized healthcare through the Internet of Things (IoT) and smart wearable devices will demand the storage and processing of progressively larger data volumes within extremely low-power systems.

Neural networks (NNs), an AI algorithm, have revolutionized various fields, offering powerful tools for pattern recognition, decision-making, and predictive analytics. However, their predictions are often overconfident, even when they are wrong or receive out-of-distribution (OOD) data. Bayesian Neural Networks (BayNNs), a subclass of NNs, overcome the crucial drawback of NNs by offering the ability to provide uncertainty estimates in predictions. The predictive uncertainty of an NN model is particularly important for critical decision-making scenarios, such as autonomous driving, medical image analysis, and industrial predictive maintenance. However, the deployment of BayNN faces significant challenges due to its inherent hardware requirements. Specifically, the 'memory wall' problem is a concern when implemented in traditional Von Neumann architectures, due to separate computing units and memory blocks. This bottleneck significantly limits the efficiency and scalability of BayNNs in practical high-demand applications.

Neuromorphic computing, emulating the brain's functions through direct neural network and synaptic array implementation, presents superior processing capabilities in AI tasks when compared to classical microprocessors. Integrating neuromorphic hardware into Green AI inherently resolves data transfer challenges by reducing reliance on cloud data transfer. This integration heralds a new paradigm poised to permeate various IT sectors and facilitate the creation of genuinely autonomous devices.


Computing-in-Memory (CIM) architectures, as a hardware realization of neuromorphic computing, have emerged as a promising solution to overcome some of these limitations. In CIM architectures, the multiply and accumulate (MAC) operation of NN can be carried out where data already exists, eliminating the memory wall problem. In particular, the integration of spintronics-based technologies into CIM offers extremely low power consumption, high endurance, and high computational efficiency~\cite{lee_world-most_2022}, while showing compatibility with existing semiconductor processes. However, such emerging advanced technologies are not without their challenges. Non-idealities like manufacturing variations and defects, as weel as stochastic behavior of spintronic memories add layers of difficulties to the implementation of BayNNs on CIM architectures.

The NeuSPIN project (\textbf{currently in the final stage}) sets out to confront the challenge of migrating cloud-based AI algorithms to On-Chip processing, emphasizing high accuracy and minimal energy consumption, primarily leveraging spintronic technology as the chosen Non-Volatile Memory (NVM) approach. Through innovative neuron and synapse designs coupled with Bayesian Neural Networks as the core AI model, the project aims to pioneer a dependable neuromorphic framework for energy-efficient AI processing at the edge. This initiative involves optimizing material stacks and device parameters specifically for AI applications, integrating spintronics technology into the fabric to minimize power consumption, and developing technology-aware statistical training algorithms. Moreover, the collaboration between the esteemed German - Karlsruhe Institute of Technology (KIT) - and the French - University of Grenoble Alpes (UGA) -  universities aims to solidify a leading partnership in Edge-AI research, positioning Europe as a front-runner in AI development at the edge.


The project is presented in the next sections. The following section~\ref{sec:Background} describes the background on spintronic memories, uncertainty estimation, BayNNs, and discusses related papers. Afterwards, in Section~\ref{sec:Challenges}, challenges associated with BayNN integration into CIM are discussed. Then, in Section~\ref{sec:Proposed} several proposed approaches are discussed. In Section~\ref{sec:Takeaways}, key takeaways of NeuSpin project are discussed. Lastly, Section~\ref{sec:Conclusion} concludes the paper.

\section{Background}\label{sec:Background}

\subsection{Spintronics Tenchnology}
The fundamental component of magnetic memory (MRAM) lies in the Magnetic Tunnel Junction (MTJ), composed of two ferromagnetic layers: the Free Layer (FL) and the Reference Layer (RL), divided by a thin insulating layer. The device's overall resistance state is determined by the relative magnetization of these layers, whether Parallel (P) or Anti-Parallel (AP).
In MRAM technology, two distinct switching mechanisms exist: Spin Transfer Torque (STT), prevalent in STT-MRAM, and Spin-Orbit Torque (SOT), employed in SOT-MRAM. Unlike STT-MRAM's two-terminal design, SOT-MRAM functions as a three-terminal device. While both offer high endurance and swift switching, SOT-MRAM stands out by segregating read and write paths within its structure. This segregation enables SOT-MRAM to provide tunable resistances, adjustable to several M$\Omega$. Such a distinctive capability holds significant promise, especially in Matrix-Vector Multiplication operations within crossbar arrays.

SOT-MRAM, comprising an MTJ mounted on a heavy metal substrate, allows also for the integration of multiple MTJs on the same layer, simulating a multi-value cell. Moreover, both STT and SOT devices exhibit stochastic switching behavior, meaning that given specific current amplitudes and durations passing through or beneath the MTJ, the switching process occurs with a certain probability. Consequently, these devices serve a dual role, functioning not just as memory storage but also potentially serving as sources of randomness.

\subsection{Uncertainty Estimation}
In NNs, uncertainty estimation is a critical process in which the network provides a measure of confidence of the prediction (predictive uncertainty) along with the prediction itself. The predictive uncertainty reflects the model's belief in its output based on the given input data and its learned parameters~\cite{nguyen2015deep}. The ability to estimate uncertainty in its prediction allows NNs to acknowledge the limits of their knowledge.

Uncertainty in deep learning is primarily divided into two types: aleatoric and epistemic. Aleatoric uncertainty, also called data uncertainty, is irreducible even if the model revives more data. This is because aleatoric uncertainty arises due to the inherent noise (randomness or variability) present in the data generation process. In contrast, epistemic uncertainty, also called model uncertainty, arises from incomplete knowledge within the model or data. This type of uncertainty can be reduced with additional data or with a more comprehensive understanding of the sources of uncertainty. The ability to estimate epistemic uncertainty empowers NNs to recognize the boundaries of their knowledge.


\subsection{Bayesian Neural Networks}
BayeNNs represent a significant shift from traditional NN paradigms during training and inference, as they embrace a probabilistic approach. Specifically, conventional NNs are described by a deterministic function $\vnet: \mathbb{R}^D \times \mathbb{R}^P \to \mathbb{R}^C$, where $D$, $P$, and $C$ represent the dimensions of inputs $\vx$, learnable parameters $\vtheta$, and outputs $\vnet(\vx, \vtheta)$, respectively. In BayNN, the parameters $\vtheta$ are not single point estimates but rather random variables, following a distribution $\vtheta \sim p(\vtheta)$. The learning objective comprises the Bayesian framework for computing the posterior distribution $p(\vtheta \mid \mathcal{D})$ given the observed data $\mathcal{D}$. 
Unfortunately, the exact computation of this posterior distribution $p(\vtheta | \mathcal{D})$ is intractable, especially when dealing with complex and complex high-dimensional models such as NNs. This is due to the high-dimensional integral calculation (over all possible values of \vtheta) in the denominator of Bayes' theorem:
\begin{align*}
p(\mathcal{D}) = \int p(\mathcal{D} | \vtheta) p(\vtheta) \, d\vtheta.
\end{align*}

As a result, direct computation of the integral in closed form becomes practically impossible, and we resort to approximation techniques such as variational inference (VI) and Monte Carlo dropout (MC-dropout). VI introduces a variational distribution $q_\vomega(\vtheta) \approx p(\vtheta \mid \mathcal{D})$, often a Gaussian with diagonal covariance, to approximate the true posterior. The parameters ($\vomega$) of this distribution are optimized to minimize the Kullback-Leibler divergence with the true posterior, leading to an efficient approximation method.

On the other hand, MC-Dropout utilizes dropout layers not only during training but, more importantly, also during inference. Thus, MC-Dropout effectively simulates samples from an approximate posterior distribution.

\subsection{Limitations and Challenges of BayNNs for CIM Implementation}\label{sec:Challenges}
Conventional BayNNs typically have either 32-bit floating-point precision~\cite{gal2016dropout} or quantized to high-bit precision, e.g., 8- to 16-bit precision~\cite{cai_vibnn_2018}, for their implications. However, since spintronic memories have only two stable states, regardless of which approximation is used for BayNNs, low-precision quantization of their parameters or learned statistical distributions is required. As a result, the quantization error is a concern for uncertainty estimates and inference accuracy.


In terms of VI, the main challenge lies in a) implementing the posterior distribution in the CIM, b) efficiently sampling from it for the inference step, as described in \cite{ahmed_spinbayes_2023}, and c) memory consumption. Specifically, on-the-fly sampling can lead to high computational costs at run time. It can be described as random sampling $\vtheta \sim q_\vomega(\vtheta)$ followed by forward passing of the given $\vx^*$ through the NN with $\vtheta$ as its parameters. 


On the other hand, for MC-Dropout, the main challenges lie in a) the realization of the dropout module in CIM, b) the sampling latency, and c) the shear number of dropout modules required. Specifically, conventional MC-Dropout applies dropout to each neuron of a layer, and another approach (MC-DropConnect) applies to each weight. Since the number of neurons and weights in an NN can be millions, the number of Dropout modules in the hardware can be huge and the overall sampling latency can be long.

The NeuSpin project, as a result, is focused on full-stack optimization, from the algorithm all the way to the circuit level, to solve the above-mentioned challenges.

\subsection{Related Works}
Previous studies have investigated hardware solutions for Bayesian Neural Networks, which are usually based on traditional CMOS technology.
In the paper~\cite{awano_bynqnet_2020}, a novel FPGA implementation is proposed that utilizes quadratic nonlinear activation functions. Work in~\cite{fan_fpga-based_2022} implemented an FPGA-based design aimed at accelerating Bayesian neural networks by employing an intermediate-layer caching technique.
The approach described in \cite{malhotra_exploiting_2020} involves a CIM implementation in which the crossbar array stores the variance parameter and stochastic resistive (RRAM) devices are used to sample the probability distribution at the input of the array. 
In~\cite{dalgaty_situ_2021}, the authors leverage the imperfections present in RRAM devices to implement Bayesian learning techniques
In~\cite{dutta_neural_2022}, the authors introduced a stochastic selector to facilitate weight dropout for neural sampling.
The research in \cite{bonnet2023bringing} showed the application of an ensemble of resistive crossbar arrays to store probabilistic weights for implementing BayNN. 
In Yang et al.~\cite{yang_all-spin_2020}, crossbar arrays were used to construct Bayesian neural networks with the help of low-barrier MTJs. The paper in~\cite{lu_algorithm-hardware_2022} presented an alternative implementation with MRAM-based crossbar arrays that can represent mean and variance. 
 
Our project aims at minimizing the reliance on Random Number Generators for Bayesian neural networks while proposing In-Memory computation to reduce excessive power consumption caused by frequent data transfers between memory and the core unit~\cite{horowitzenergy}. To achieve this, we adopt a hardware-software co-design approach, proposing architectural solutions involving fewer stochastic units and CIM architecture. 
Moreover, we leverage all spintronics functionalities in our hardware implementation to achieve a low-power design. Overall, the proposed approaches mitigate the inherent stochasticity of the spintronic devices by considering them as a feature rather than a foe.

\section{Scalable Spintronic-Based Bayesian NNs}\label{sec:Proposed}
As mentioned, BayNNs are the principal methods for estimating predictive uncertainty in a machine learning application. Despite this, the computational cost and power consumption make the use of BayNNs on
embedded hardware (with limited resources) is unattractive. Conventional BayNNs suffer from the \emph{resource scalability} issue. As the size of the models and datasets increases, the computational and memory requirements also increase. In the case of BayNN implementations, this also means that the number of Random Number Generators (RNGs), to be realized by stochastic spintronic devices, and memory consumption to store model parameters also increases with model size. For example, in MC-Dropout~\cite{gal2016dropout}  and MC-DropConnect~\cite{mobiny2021dropconnect}, the number of Dropout modules equals the total number of weights and activations in the model. Similarly, the memory consumption of certain VI and ensemble implementations can be $2 -10\times$ higher. This means that achieving the desired performance with minimal energy consumption, memory usage, and latency is challenging for BayNNs. Efficient resource scalability is particularly important in resource-constrained environments with edge AI, such as mobile devices or embedded systems. Scalable approaches for BayNN realization based on spintronic technology, as a part of the NeuSpin project, are described below.



\subsection{Dropout Based BayNN}
MC-Dropout typically has a lower memory consumption compared to other approximation methods, as it has the same number of point estimation parameters as a conventional NN. However, they have their own challenges, as mentioned earlier. We present the dropout-based BayNN approaches considered in the NeuSpin project in the next sections.

\subsubsection{SpinDrop}
In this project, we introduce for \textbf{the first time the binary Bayesian Neural Network (BinBayNN)}\cite{soyed_nanoarch22,soyed_spindrop} utilizing the inherent stochastic properties of spintronic devices for the implementation of the dropout module and the deterministic properties of BayNN weight storage for in-memory Bayesian inference.
To implement dropout-based BayNN, each neuron in the array was equipped with a dedicated dropout module, which allowed the potential dropout of a neuron based on a specified probability.
To enable control over the current and, consequently, the probability of the MTJ, CMOS transistors were integrated with the MTJ. The process involved generating a bitstream by alternating SET and RESET operations. Following a "SET" write operation, the MTJ's state was read using a sense amplifier to verify the occurrence of the switch, effectively indicating the dropout signal. Post-read operation, the MTJ was "RESET" to the P-state in preparation for the next round of dropout signal generation~\cite{soyed_spindrop, soyed_nanoarch22}.  In classic BNNs, the standard matrix-vector multiplications are replaced with XNOR operations. Thus, it is necessary to design the spintronic-based bit cell that allows for this particular operation. The proposed architecture is organized around the MTJ crossbar array, in which each trained weight is stored in a
unit represented by two 1T-1MTJ cells. In addition, a word-line decoder is used with the capability to enable multiple consecutive addresses. A SpinDrop module, described in the previous section, is
connected to each pair of word-line to implement the dropout concept.

A specifically designed learning objective is proposed. Evaluation of the proposed approach shows up to 100\% detection of out-of-distribution data, an improvement in accuracy of $\sim 2\%$, and up to $15\%$ for corrupted data.

\subsubsection{Spatial-SpinDrop}
\begin{figure}
\centering
   \includegraphics[width=0.8\columnwidth]{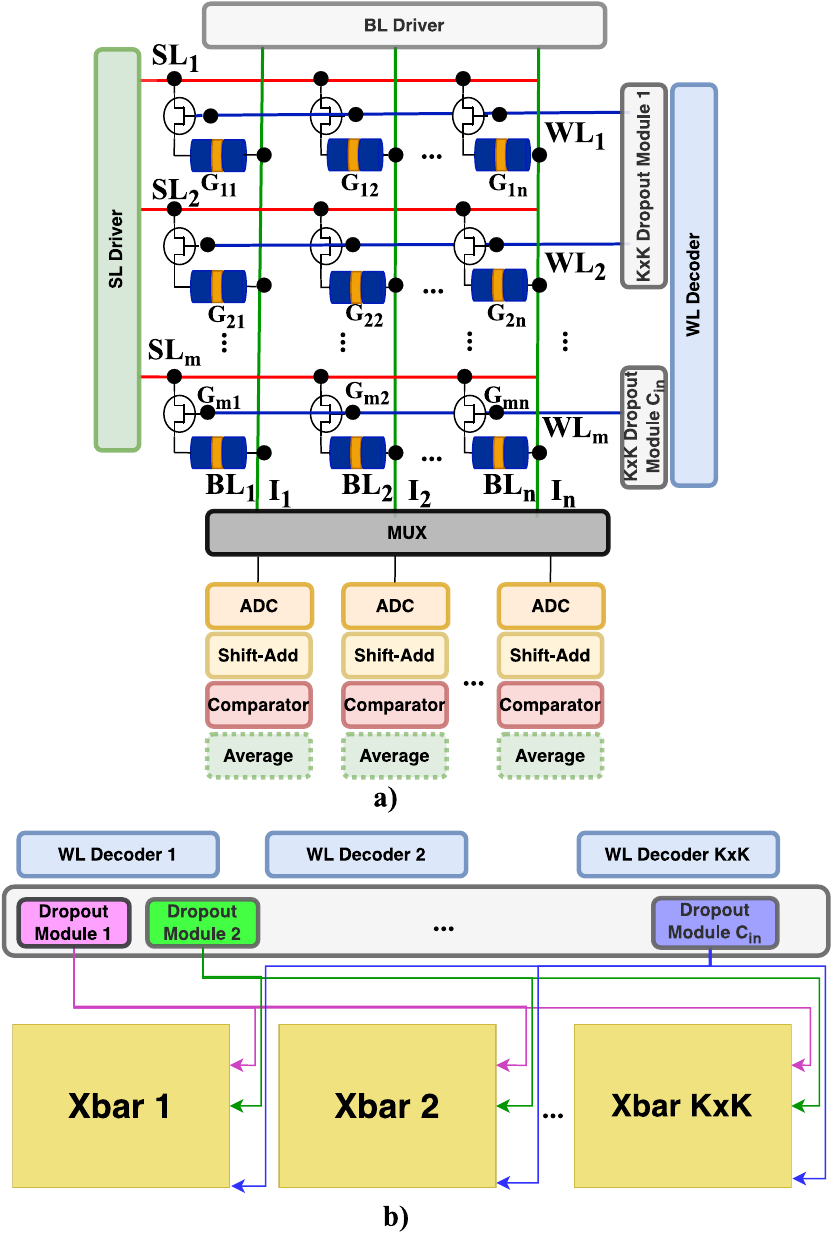}
\vspace{-0.5\baselineskip}   
  \caption{Crossbar design for the MC-SpatialDropout based on mapping strategy (a)  \textcircled{1} and (b) strategy\textcircled{2}. In (b), only the Dropout module and WL decoder are shown. Everything else is abstracted. }
\label{fig:crossbar}
\end{figure}
However, one drawback of the SpinDrop structure is that it requires a considerable number of dropout modules per layer, typically one for each neuron. Although the dropout module can be reused, the number of dropout modules still increases with network size. Also, such an implementation is not adapted for convolutional layers.

Expanding on the SpinDrop concept, we developed the MC-SpatialDropout based~\cite{soyed_TNANO23} BayNN to reduce the number of dropout modules and allow easier implementation of the dropout module for convolutional layers. In MC-SpatialDropout, instead of the conventional dropout module that randomly disables neurons, spatial dropout is used. Spatial dropout drops entire feature maps, making it more suitable for CNNs where spatial correlations are vital. Furthermore, this method reduces the complexity of circuit design for implementing dropout in CNNs. The overall effort to co-design the algorithm and hardware leads to a reduction in the number of dropout modules per network by a factor of $9\times$ and energy consumption by $94.11\times$, while maintaining comparable predictive performance and uncertainty estimates, the architecture is depicted in Fig.\ref{fig:crossbar}.

In addition, two prevalent mapping strategies are explored for mapping the convolutional layer with SpatialSpinDrop. In the first strategy (\textcircled{1}), every kernel, shaped as $K\times K\times C_{in}$, is unfolded into a crossbar column, as outlined in~\cite{gokmen_training_2017}.On the contrary, the second strategy (\textcircled{2}) involves mapping each kernel onto smaller crossbars of $K\times K$ dimensions, with a shape of $C_{in}\times C_{out}$, as detailed in~\cite{peng_optimizing_2019}. As a result, the Dropout module needs to be generalizable to different mapping strategies, which is not the case with the SpinDrop concept. Thus, the MC-SpatialDropout approach is $2.94\times$ more energy efficient than the SpinDrop concept presented earlier.


\subsubsection{SpinScaleDrop}
\begin{figure}
   \centering
   \includegraphics[width=0.75\columnwidth]{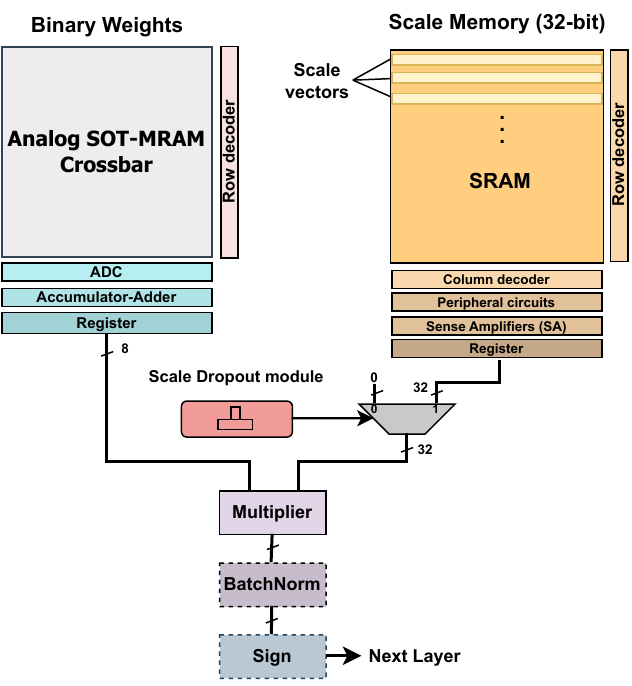}
   \caption{Proposed inference architecture for Scale-Dropout.}
   \label{fig:architecture}
\end{figure}

To further improve the efficiency of BayNN, we introduced a novel dropout approach called SpinScaleDrop \cite{ahmed2023scale}. In the scale dropout approach, a \emph{scalar dropout mask} is applied to the scale vector by scale modulation rather than information zeroing for each layer. Thus, \textbf{only a single dropout module is per layer}. The scale vector is defined as a learnable parameter through gradient descent, and its shape is specifically designed to be suitable for the CIM architecture and to have low memory overhead. SpinScaleDropout introduces randomness in the scale vector and, thus, the activation of a layer, potentially reducing co-adaptation between the scale vector and binary weights. 

Furthermore, a \emph{layer-dependent adaptive scale dropout} approach is proposed that selects the dropout probability based on the parameter size of the layer. Thus, extensive design-space exploration is mitigated to find the optimal dropout probability and location of the dropout layer for each model. The proposed BayNN with Scale Dropout can be trained using stochastic gradient descent. The training process involves sampling a scaled network for each forward pass, and the gradients for each parameter are averaged over the training instances of each mini-batch. In terms of using scale-dropout as a Bayesian approximation, the learning objective involves a novel regularization function for the scale vector to encourage it to be positive and centered around one. This regularization approach aligns well with the nature of binary NN, where the weights are binarized to $\pm 1$.

Model uncertainty can be obtained by performing multiple forward passes with an enabled scale dropout during Bayesian inference. Each forward pass generates a stochastic scale vector, allowing Monte Carlo estimates of the model’s predictive mean and uncertainty.

In terms of hardware implementation, spintronics-based scale dropout is modeled differently due to the manufacturing and in-field variation of the MTJs in the spintronic-based Scale-Drop module. Specifically, the dropout probability is defined as a stochastic variable, and the dropout probability is fitted to a Gaussian distribution. Such a module allows for the stochastic activation of the Scale vector that is stored in a neighboring memory. 
Furthermore, a novel CIM architecture is proposed that incorporates a SOT-MRAM crossbar array, an SRAM memory for scale vector storage, and a spintronics-based scale dropout module for stochastic activation, see Fig.\ref{fig:architecture}.

In terms of results, we have demonstrated up to $1\%$ improvement in predictive performance. Furthermore, our approach showed more than $100\times$ energy savings compared to existing methods.

\subsubsection{Inverted Normalization with Affine Dropout}

The approach introduces a novel normalization and dropout method to enhance the robustness and inference accuracy of BayNNs, \textbf{a self-healing BayNN}, particularly in CIM architectures~\cite{ahmed2024inverted}.

In traditional normalization techniques, such as batch or layer normalization, normalization is first performed, and then an optional affine transformation is performed. Unlike the traditional approach, the inverted normalization layer treats the affine parameters ($\beta$ and $\gamma$) as similar to the weights and biases of the NN, that is, their primary objective is to minimize loss via gradient descent. Furthermore, the affine transformation is \emph{necessary} and performed before normalization, ensuring that the learning process remains stable.

Furthermore, the Affine Dropout approach is introduced, which adds stochasticity to the weighted sum by randomly dropping the weight and bias of the inverted normalization layer with probability $p$. The implementation involves sampling two binary dropout masks, one for weight and the other for bias of Affine Dropout, from a Bernoulli distribution. Dropout masks are kept at a scalar value (vector-wise dropout) instead of a vector (element-wise dropout) to reduce the number of RNGs in the model. The sampled masks are then multiplied by the weights and biases of the inverted normalization layer, setting the dropped weights and biases to ones and zeros, respectively. 

In terms of Bayesian inference and uncertainty estimation, the proposed affine dropout acts as an alternative to the conventional dropout in a Bayesian setting. Thus, during Bayesian inference, multiple forward passes through the network with independently sampled weights and bias masks for each layer result in a stochastic output distribution. The output distribution is treated as an approximation of the Gaussian process, following the theoretical foundation of existing work~\cite{gal2016dropout}. 

The evaluation of the approach shows an improvement in inference accuracy by up to 55.62\%. In addition, in specific models such as LSTM-based time series prediction, the root mean square error (RMSE) score is reduced by up to 46.7\%. Furthermore, the proposed method is capable of detecting up to 55.03\% and 78.95\% of OOD instances for uniform noise and random rotation experiments, respectively. As a result, an improvement of 14.61\% is achieved compared to previous work.

\subsection{Variational Inference (VI) Based}
In terms of inference accuracy and quality of uncertainty, Variational Inference (VI) typically offers a more precise inference and a more accurate representation of the uncertainty. However, the challenges lie in their implementation in the CIM architecture as a result of their distributional parameters and on-the-fly sampling requirements. In this section, we briefly discuss the full-stack optimization approaches for VI-based BayNN from NeuSpin.

\subsubsection{Bayesian Sub-Set Parameter Inference}

In conventional VI-based BayNNs, the variational distributions are typically applied to weights or neuron activation.
Consequently, memory consumption and computational complexity are high, as weights represent a large proportion (more than $90\%$) of memory utilization. Furthermore, implementation challenges in CIM-based hardware due to limited stable states are a factor, as VI-based BayNNs are not suitable for binary weights.

In this project, we introduce \emph{Bayesian subset parameter inference}, an efficient and scalable Bayesian NN framework with both deterministic and stochastic parameters\cite{ahmed2023scalable}. Larger parameter groups (e.g., weights) are kept deterministic, while Bayesian treatment is only applied to the small parameter group, e.g., scale vector. Specifically, Bayesian treatment is applied to learn the distribution of the scale vector, while other parameters are learned via maximum likelihood. Consequently, our approach is the \textbf{first binary VI-based BayNNs framework with spintronic-based CIM implementation}.

We proposed a novel CIM architecture with two separate crossbars per layer: one for deterministic weights and the other for the Bayesian scale. Furthermore, a multi-level device composed of multiple MTJs is implemented to quantitatively represent Bayesian parameters. Additionally, the stochastic behavior of the SOT devices is utilized as a random number generator for sampling.

Experimental results show that the proposed approach has comparable accuracy to full-precision models while estimating uncertainty efficiently. The effectiveness in handling out-of-distribution data is demonstrated by the increase in negative log-likelihood (NLL) under dataset shifts. Furthermore, the proposed approach requires up to $70\times$ lower power consumption and $158.7\times$ lower storage memory requirements compared to traditional methods.

\subsubsection{SpinBayes}

\begin{figure}
    \centering
    \includegraphics[width=0.7\columnwidth]{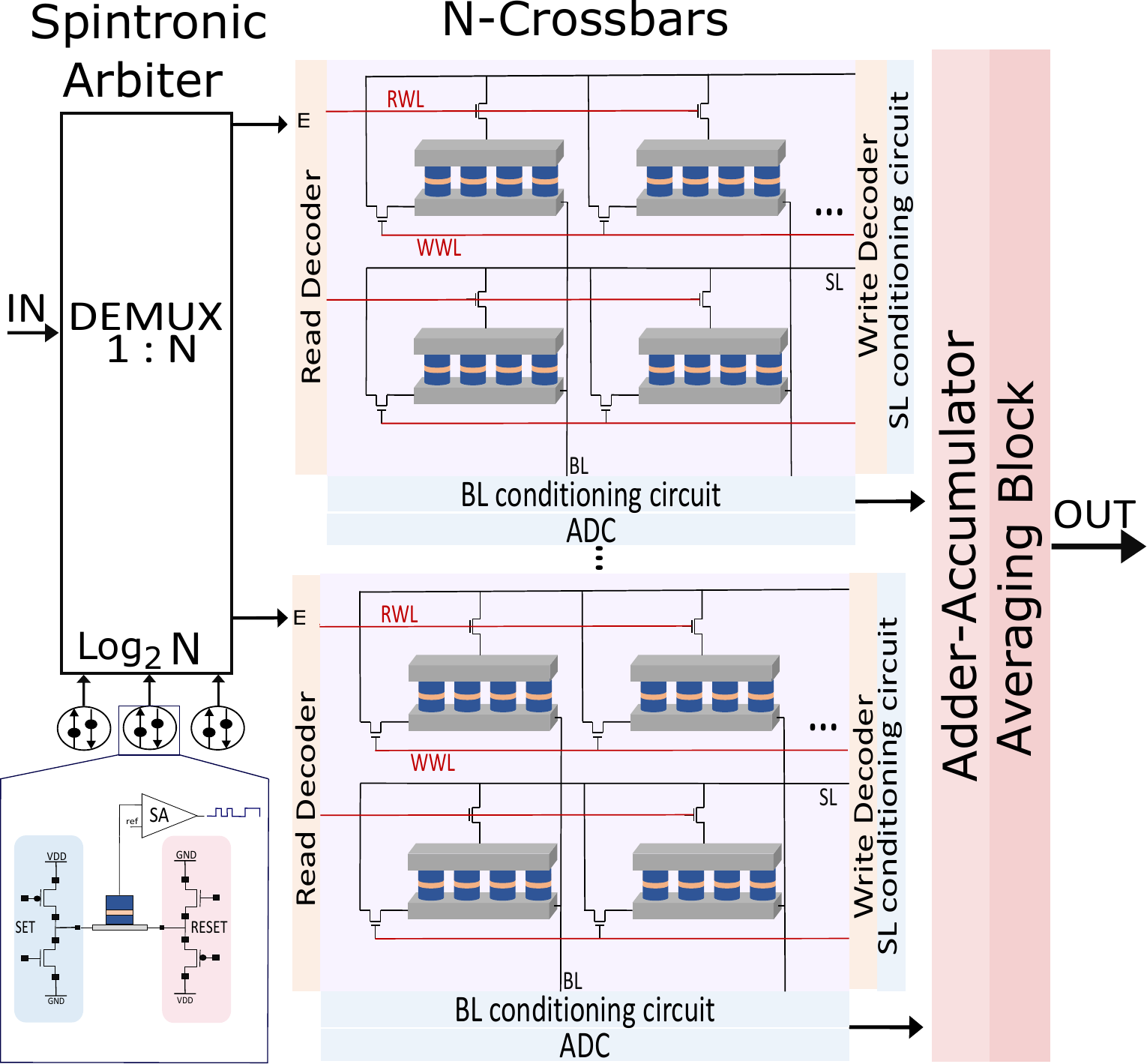}
    \caption{Proposed layer architecture}
    \label{fig:Architecture_ensemble}
\end{figure}

In this project~\cite{ahmed_spinbayes_2023}, we introduce \emph{Bayesian in-memory approximation} for efficiently mapping the posterior distribution of BayNN to spintronic-based CIM architecture. The main idea behind the approach is to create a memory-friendly distribution that can be efficiently mapped and sampled by CIM hardware. Furthermore, to enable efficient sampling from the mapped posterior distribution on CIM hardware, a novel network topology named "SpinBayes" is introduced. 

The SpinBayes topology presented in Fig.~\ref{fig:Architecture_ensemble} is tailored for CIM architectures leveraging spintronics and incorporates an Arbiter for each layer. This spintronic structure is built upon a multi-value SOT bit-cell crossbar array. Within a layer, the spintronic stochastic Arbiter is implemented at the periphery of crossbars, selecting specific crossbars for Bayesian inference in each forward pass. The Arbiter generates a random binary one-hot vector to determine the selection.

Furthermore, the proposed approximation approach introduces CIM- and spintronics-aware post-training quantization. In terms of the memory cells in the crossbar, we proposed novel MTJ-based multi-value cells for quantized weight storage. In our proposed design, a certain number of conductance levels can be programmed. We perform design-time exploration to optimize bit-precision, followed by MTJ-based multi-value design.

The proposed method is evaluated in classification tasks with up to $100$ classes and semantic segmentation tasks on two safety-critical tasks: medical image diagnosis and automotive scene understanding tasks. Compared to SOTA methods, improvements in classification accuracy of up to $1.14\%$ and uncertainty estimation of up to $20.16\%$ are shown. Furthermore, the proposed approach can detect up to $100\%$ samples from several out-of-distribution datasets.
\begin{table*}[]
\caption{Comparison of Methods}
\centering
\begin{tabular}{llll}
\cline{2-4}
\multicolumn{1}{l|}{}                             & \multicolumn{1}{l|}{\textbf{Method}}                     & \multicolumn{1}{l|}{\textbf{Inference accuracy}} & \multicolumn{1}{l|}{\textbf{Energy}}        \\ \hline
\multicolumn{1}{|l|}{\textbf{Dropout Based}}               & \multicolumn{1}{l|}{SpinDrop}                   & \multicolumn{1}{l|}{91.95\%}                         & \multicolumn{1}{l|}{2.00 $\mu$ J/Image} \\ \cline{2-4} 
\multicolumn{1}{|l|}{}                            & \multicolumn{1}{l|}{Spatial-SpinDrop}           & \multicolumn{1}{l|}{90.34\%}                         & \multicolumn{1}{l|}{0.68 $\mu$ J/Image} \\ \cline{2-4} 
\multicolumn{1}{|l|}{}                            & \multicolumn{1}{l|}{SpinScaleDropout}           & \multicolumn{1}{l|}{90.45\%}                         & \multicolumn{1}{l|}{0.18 $\mu$ J/Image} \\ \hline
\multicolumn{1}{|l|}{\textbf{Variational Inference Based}} & \multicolumn{1}{l|}{Bayesian Sub-Set Parameter} & \multicolumn{1}{l|}{90.62\%}                         & \multicolumn{1}{l|}{0.30 $\mu$ J/Image} \\ \cline{2-4} 
\multicolumn{1}{|l|}{}                            & \multicolumn{1}{l|}{SpinBayes}                  & \multicolumn{1}{l|}{--}                              & \multicolumn{1}{l|}{0.26 $\mu$ J/Image} \\ \hline
                                                  &                                                 &                                                      &                                    \\
                                                  &                                                 &                                                      &                                   
\end{tabular}
\vspace{-3em}
\end{table*}

\section{Key Takeaways and Future Works}\label{sec:Takeaways}
The key takeaways from the NeuSpin project can be summarized as \begin{enumerate*}
    \item \textbf{Effective Detection of Out-of-Distribution Data}: In multiple studies, we find that BayNNs have shown the ability to detect up to $100\%$ out-of-distribution data, highlighting their robustness in handling unexpected or noisy input.
    \item \textbf{Improvement in Inference Accuracy for Corrupted Data}: We find that the implementation of Bayesian methods has a significant improvement in inference accuracy, especially in scenarios involving corrupted data. This improvement is critical in applications where data integrity cannot always be guaranteed.
    \item \textbf{Incorporation of Spintronic Technologies}: The use of spintronic-based memories, particularly in CIM architectures, is advantageous in terms of lower energy consumption and switching speed, which are essential for efficient hardware implementations of BNNs. Also, by using their stochastic regime, RNG implementation is possible using the same technology.
    \item \textbf{Modeling defects in Devices}: Understanding and accounting for modeling defaults in devices is crucial. These defaults might include inherent variations, imperfections, or manufacturing irregularities in the spintronic-based devices used for BayNN implementations. Incorporating these defaults into the modeling process helps in creating more accurate representations of the hardware and leads to better algorithm-hardware co-design strategies.
    
    \item \textbf{Algorithm-Hardware Co-Design}: The trend towards algorithm-hardware co-design, where Bayesian algorithms are specifically tailored to leverage the features of modern hardware architectures like spintronics and CIM, enables more efficient and practical implementations of BayNNs.
    \item \textbf{Quantized BayNNs}: Quantization or binarization of parameters and activations in BayNNs is a key strategy not only to reduce memory overhead but also to overcome limited states of spintronics memories. Thus, some of the inherent costs of BayNNs can be reduced, making them more feasible for deployment in resource-constrained safety-critical applications.
    \item \textbf{Resource-aware Methods for Bayesian Approximation}: Novel dropout techniques, such as Scale Dropout and Affine Dropout, and memory-centric approximations allow not only the reduction of the number of RNGs but also the efficient implementation of BayNN. 
    \item \textbf{Inherent Robustness and Self-healing}: BayNNs are inherently robust to variations and faults compared to conventional NNs. Their self-healing property can be further enhanced with specifically designed techniques such as inverted normalization with Affine Dropout.
\end{enumerate*}


\section{Conclusion}\label{sec:Conclusion}
Bayesian NNs are inherently capable of estimating uncertainty in predictions, overcoming the limitations of conventional NNs. However, they are inherently resource-hungry in terms of power, memory consumption, latency, and the number of RNGs per model. Some of their  costs can be reduced by implementing them in a spintronics-based CIM architecture. However, implementing BayNN in a CIM architecture is not straightforward and has several challenges. The objective of NeuSpin project, as a French-German collaborative project, is full stack optimization, from algorithm, BayNN topology, to circuit design, to optimize BayNN training, inference, and CIM implementation using spintronic non-volatile technology. This can greatly reduce hardware footprint while providing energy efficiency and uncertainty estimation for critical edge AI application. Key achievements include the ability to detect up to 100\% of out-of-distribution data, consistently improve inference accuracy, especially in the presence of corrupted data, and enhance robustness against non-ideal properties of spintronics. Additionally, a notable breakthrough is achieved with the integration of algorithmic innovation with advanced hardware technologies such as spintronics and CIM architectures, leading to substantial reductions in energy consumption (up to $100\times$) and memory overhead ($158.7\times$). Consequently, our novel approaches have paved the way for efficient, reliable, and scalable BayNNs in various real-world and safety-critical applications. The collaboration between the two partners has been a success so far, thanks to the  complementary competences ans skills of the two teams, but also thanks to the possibility to use realistic data and parameters of spontronic devices fabricated and characterized in the SPINTEC facility.

\bibliographystyle{IEEEtran}
\typeout{}
\bibliography{references}

\end{document}